\documentclass[reprint,amsmath,amssymb, aps, prb,
floatfix,nofootinbib]{revtex4-1} 
\usepackage{graphicx}
\usepackage{dcolumn}
\usepackage{bm}
\usepackage{caption}
\usepackage{amssymb,amsmath,color}
\usepackage{url}
\usepackage{float}
\usepackage{braket}
\usepackage{hyperref} 
\usepackage{comment}
\hypersetup{colorlinks = true, linkcolor = blue, anchorcolor = blue, citecolor = blue, filecolor = blue, urlcolor = blue}
\usepackage{xcolor}
\usepackage{empheq}
\usepackage{multirow}

\begin{document}
\title{Exciton multipolarity controls coherent and squeezed phonons in van der Waals heterostructures}
\author{
  Indrajit Maity\textsuperscript{1,*}, 
  Arash A. Mostofi\textsuperscript{2},
  Johannes Lischner\textsuperscript{2},
  \'{A}ngel Rubio\textsuperscript{3,4} \\ [1ex]
  \textsuperscript{1}Department of Chemistry, School of Natural and Environmental Sciences, Newcastle University, Newcastle upon Tyne, NE1 7RU, UK.\\  
  \textsuperscript{2}Departments of Materials and Physics and the Thomas Young Centre for Theory and Simulation of Materials, Imperial College London, South Kensington Campus,
London, SW7 2AZ, UK. \\
  \textsuperscript{3}Max Planck Institute for the Structure and Dynamics of Matter, Luruper Chaussee 149, 22761 Hamburg, Germany. \\
  \textsuperscript{4}Initiative for Computational Catalysis and Center for Computational Quantum Physics, Flatiron Institute, Simons Foundation, New York City, New York 10010, USA.\\ 
  \textsuperscript{*}Corresponding author: indrajit.maity@newcastle.ac.uk
}
\date{\today}

\begin{abstract}
Photoexcitation-driven changes in the electronic distribution displace atoms, generating coherent phonons on ultrafast timescales. Two-dimensional (2D) materials and their heterostructures offer a powerful platform for engineering these phonons. Yet, despite the widespread observation of photoexcited coherent phonons, a design principle for controlling their character remains elusive. Here, using detailed atomistic simulations of multilayers of alternating MoSe$_2$ and WSe$_2$, we reveal exciton multipolarity as a design principle for tuning photoinduced phonons from coherent to squeezed. These phonons are interlayer breathing modes, with dipolar excitons coupling linearly to generate coherent states and quadrupolar excitons coupling quadratically to produce squeezed states. Moreover, an out-of-plane electric field enables switch-like control, converting quadrupolar excitons into dipolar excitons and switching the phonon state from squeezed to coherent. For example, in trilayer WSe$_2$/MoSe$_2$/WSe$_2$, the photoexcited 1.04-THz breathing mode switches from a squeezed state at zero field to a coherent state under an applied vertical field. Experimentally, these phonon states can be directly probed by ultrafast X-ray or electron diffraction and indirectly through transient reflectivity. Our results open new avenues for ultrafast control of lattice and electronic dynamics on picosecond timescales, with implications for THz quantum phononics, nanophotonic technologies, and quantum-noise-limited sensing. 
\end{abstract}

\maketitle

Ultrafast processes in materials unfold on extremely short timescales, ranging from femtoseconds to picoseconds (ps). Tremendous advances in optical and X-ray techniques now resolve these processes in real time. In pump–probe spectroscopy, for example, a pump pulse drives the material out of equilibrium, while a time-delayed probe pulse captures snapshots of its evolving response~\cite{Zewailfemtochemistry2000,Claudioultrafast2016}. These advances have transformed our understanding of how photoexcited electrons couple to phonons and drive structural changes. For example, femtosecond X-ray diffraction in bismuth revealed that photoexcitation displaces atoms along the $A_{1g}$ phonon mode, softens the bonds, and drives the crystal towards a higher-symmetry structure~\cite{Fritzultrafast2007,Sokolowskifemtosecond2003}. Beyond fundamental insight, understanding and controlling these ultrafast processes will enable faster optoelectronic switches~\cite{Sieultrafast2019}, more efficient solar cells~\cite{Paulhot2021,Lincarrier2024}, and light-driven information processing at unprecedented petahertz speeds~\cite{Heidepetahertz2024,Boolakeelight2022}.

A key manifestation of photoexcited electron-phonon coupling is the generation of \textit{coherent} phonons~\cite{Zeigertheory1992,Kuznetsovtheory1994}, similar to coherent states in quantum optics~\cite{Gerryintroductory2004}. These phonons show an oscillating average displacement while their fluctuations remain stationary in time. 2D semiconducting materials and their van der Waals heterostructures provide unique platforms to realize and engineer these coherent phonons~\cite{Niecoherent2025}. Their large electron–hole binding energies lead to strongly bound excitons following photoexcitation. As a result, coherent phonons couple to excitons rather than independent electrons and holes, giving rise to exciton-dependent dynamics~\cite{Trovatellostrongly2020,Jeongcoherent2016,Sayersstrong2023,Perfettotheory2024}. Previous studies have engineered these modes by using van der Waals heterobilayers to tune electronic structure~\cite{Kimultrafast2024,Licoherent2023} or by increasing the layer number of the same material to tune phonons~\cite{Jeongcoherent2016}. For example, photoexcitation of a type-II heterojunction in MoSe$_2$/WSe$_2$ drives a breathing mode (BM) at 0.8 THz, where the two layers move against each other along the out-of-plane direction. On the other hand, photoexcitation of multilayer WSe$_2$ drives the lowest-energy BM, whose frequency is highly sensitive to the number of layers. Beyond 2D semiconductors, coherent phonons have also been realized in bulk semimetals~\cite{Zeigertheory1992,Emeiscoherent2025}, superconductors~\cite{Yangmode2019}, and topological insulators~\cite{Yanglight2020}, pointing to broader potential for realizing and controlling coherent phonons across the 2D analogs of these material families. However, existing approaches tune electrons or phonons independently, modifying only the amplitude or frequency of the resulting coherent phonons. No design principle exists for controlling the underlying character of the phonon state itself.

\begin{figure*}[ht!]
    \centering
    \includegraphics[width=1.0\linewidth]{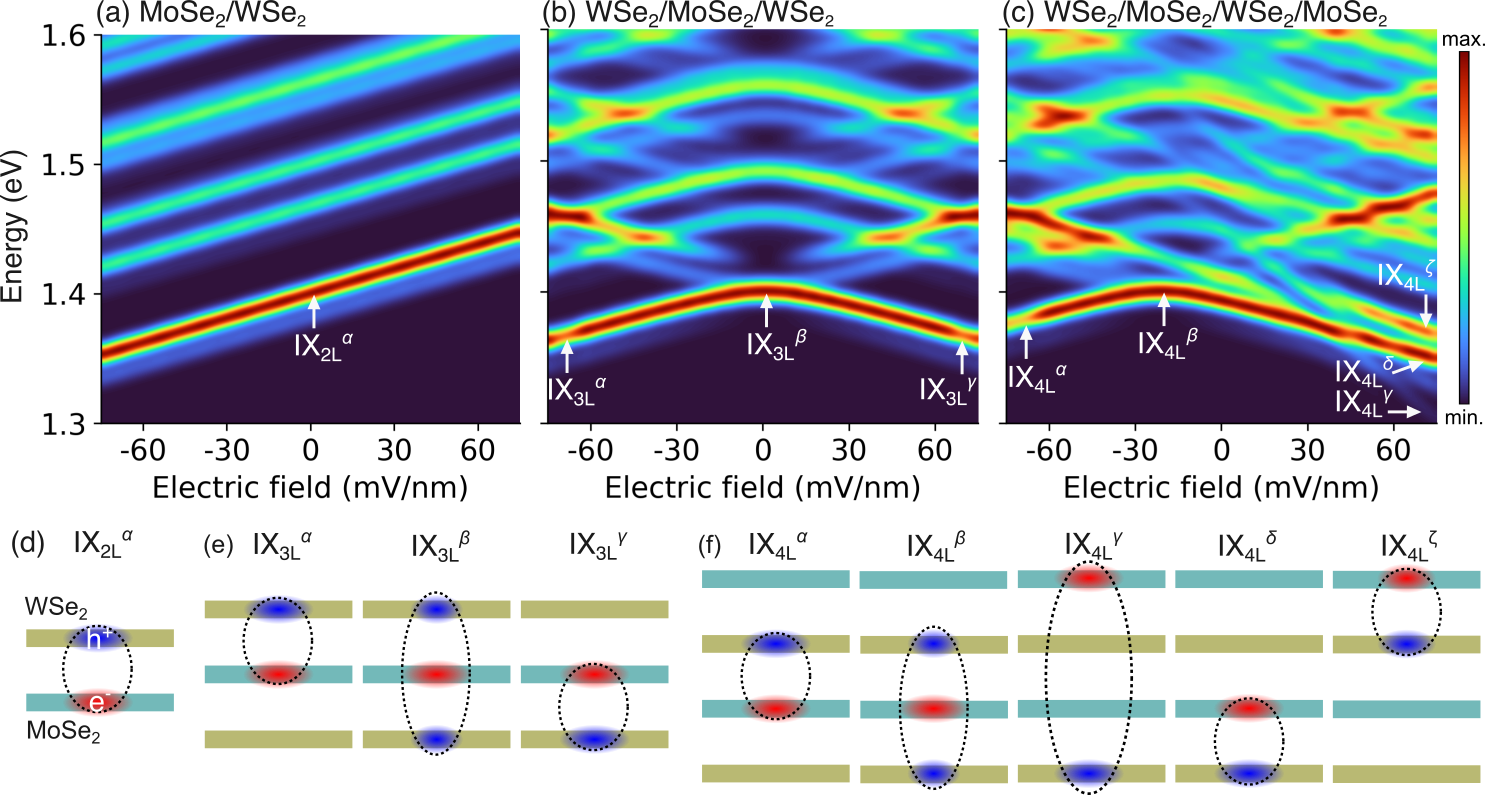}
    \caption{
    \textbf{Out-of-plane electric-field response of excitons in multilayers of MoSe$_{2}$ and WSe$_{2}$.} (a) The bilayer shows a linear Stark effect across the entire electric-field range, (b) the trilayer shows a nonlinear Stark response near zero field that becomes linear at high fields, and (c) the tetralayer shows a more complex response with both linear and nonlinear Stark shifts at finite fields, and asymmetry around zero field. (d)-(f) Schematics of the interlayer excitons (IX) at different electric fields, showing the electron and hole distributions. Each slab represents a MoSe$_{2}$ or WSe$_{2}$ layer. The colorbar denotes optical conductivity.
    }
    \label{fig1}
\end{figure*}

Here, we establish exciton multipolarity—whether the exciton is dipolar or quadrupolar—as a design principle in semiconducting van der Waals heterostructures for engineering quantum phonon states, enabling the BMs to be driven into either a \textit{coherent} or a \textit{squeezed} state. Unlike coherent phonons, squeezed phonons have zero average displacement while their fluctuations oscillate in time. Using multilayers of MoSe$_2$ and WSe$_2$ as representative examples, we show through fully atomistic exciton-phonon coupling calculations that the coherent-to-squeezed crossover arises from a linear-to-quadratic transition in the exciton-breathing-mode coupling. In bilayer MoSe$_2$/WSe$_2$, the lowest-energy exciton is dipolar and couples linearly to the only BM at 0.82 THz, producing coherent phonons consistent with experiment~\cite{Licoherent2023}. By contrast, in trilayer WSe$_{2}$/MoSe$_2$/WSe$_2$, the lowest-energy exciton is quadrupolar and couples quadratically to the highest-energy BM at 1.04 THz, producing squeezed phonons. An external vertical electric field transforms the quadrupolar exciton into a dipolar one, switching the coupling from quadratic to linear and the phonon state from squeezed to coherent. Our estimated coherent-phonon amplitudes and squeezing parameters are comparable to those in well-established bulk systems. The same design principle extends to thicker multilayers, offering a general route to electrically control quantum phonon states.

\begin{figure*}[ht!]
    \centering
    \includegraphics[width=1.0\linewidth]{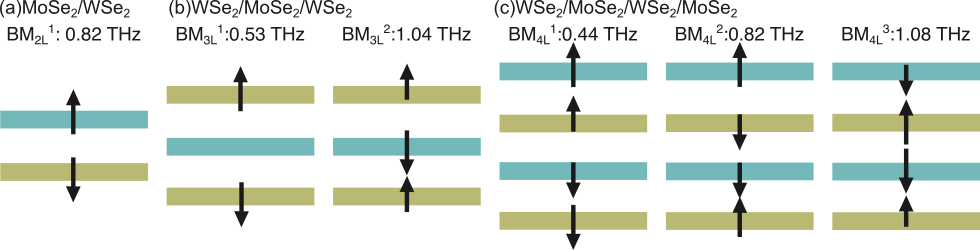}
    \caption{
    \textbf{Breathing phonons in multilayers of MoSe$_{2}$ and WSe$_{2}$.} Breathing-mode eigenvectors for a bilayer (a), trilayer (b), and tetralayer (c), with arrows indicating the out-of-plane motion of each layer. Each slab represents a MoSe$_2$ or WSe$_2$ layer.
    }
    \label{fig2}
\end{figure*}

\section{Results}
\subsection{Excitons in multilayers of MoSe$_{2}$ and WSe$_{2}$} 
We consider multilayer structures consisting of up to four layers, with alternating WSe$_2$ and MoSe$_2$ along the out-of-plane direction. Neighboring layers have 2H stacking, with metal atoms above chalcogen atoms and chalcogen atoms above metal atoms. We use a lattice constant of 3.32~\AA\ for both MoSe$_{2}$ and WSe$_{2}$, as the lattice mismatch is below 0.2\%~\cite{Brixnerpreparation1962}. Fig.~\ref{fig1} shows the optical conductivities of the multilayers under different out-of-plane electric fields, $E_z$, calculated using the spinor Bethe-Salpeter equation (spinor-BSE), which employs two-component spinor wavefunctions to distinguish bright singlet and dark triplet excitons. Throughout, we focus on the singlet states. The applied field induces Stark shifts in the lowest-lying exciton energies, with the Stark response becoming increasingly complex as more layers are added. In bilayer MoSe$_2$/WSe$_2$, the Stark response is linear for all applied $E_z$ (see Fig.~\ref{fig1}(a)). In contrast, the symmetric trilayer WSe$_2$/MoSe$_2$/WSe$_2$ deviates from this behavior, showing a nonlinear response near zero field and a transition to a linear response at higher fields (see Fig.~\ref{fig1}(b)). The tetralayer MoSe$_2$/WSe$_2$/MoSe$_2$/WSe$_2$ shows a more complex response, with a nonlinear behavior near $-21$~mV/nm and an asymmetry around zero field that is absent in the trilayer (see Fig.~\ref{fig1}(c)).

To uncover the origin of the complex Stark response in multilayers, we analyze the wavefunctions of the lowest-energy excitons at several electric fields. Fig.~\ref{fig1}(d)-(f) shows the corresponding electron and hole distributions obtained from our BSE calculations. In all multilayers, the lowest-energy excitons are interlayer excitons (IXs). The electron is localized in MoSe$_2$ and the hole in WSe$_2$. The bilayer IX hosts a permanent dipole (see Fig.~\ref{fig1}(d)). Unlike the bilayer, the trilayer IX changes its multipolar character with field, evolving from an upward dipole at $E_z=-60$~mV/nm to a quadrupole at $E_z=0$, and finally to a downward dipole at $E_z=+60$~mV/nm (see Fig.~\ref{fig1}(e)). The tetralayer IX shows an even richer evolution (see Fig.~\ref{fig1}(f)). It forms an upward dipole between the inner MoSe$_2$ and WSe$_2$ layers at $E_z=-60$~mV/nm, a quadrupole at $E_z=-21$~mV/nm with the outer MoSe$_2$ layer remaining largely inactive, and a downward dipole between the outermost MoSe$_2$ and WSe$_2$ layers at higher fields. Beyond the lowest IX, the tetralayer hosts two additional downward dipolar configurations at higher energy (see Fig.~\ref{fig1}(f)). All distinct multipole configurations are essential to understanding the linear and nonlinear Stark responses across the multilayers.

To explain the field-driven crossover from nonlinear to linear Stark response, we use model Hamiltonians. The models account for all relevant dipolar configurations, with electrons and holes tunneling between layers with amplitudes $t_e$ and $t_h$, respectively. We do not couple configurations connected only by a simultaneous electron and hole hop. The corresponding Hamiltonians are
\begin{subequations}
\label{modelH}
\begin{align}
H_{\mathrm{2L}} &= \mathcal{E}_{d} + p_{0}E_z,\\[4pt]
H_{\mathrm{3L}} &=
\begin{pmatrix}
\mathcal{E}_d + p_{0}E_z & -t_h \\
-t_h & \mathcal{E}_d - p_{0}E_z
\end{pmatrix},\\
H_{\mathrm{4L}} 
&= \begin{pmatrix}
\mathcal{E}_d^{1}-p_{0}E_z & -t_h & 0 & -t_e\\
-t_h & \mathcal{E}_d^{2}+p_0E_z & -t_e & 0\\
0 & -t_e & \mathcal{E}_d^{3}-p_{0}E_z & -t_h\\
-t_e & 0 & -t_h & \mathcal{E}_d^{4}- 3p_0E_z
\end{pmatrix}
\end{align}.
\label{avoided_crossing}
\end{subequations}
Here, $\mathcal{E}_{d}^{1-3}$ and $\mathcal{E}_{d}^{4}$ denote the energies of dipolar excitons with dipole moments $p_0=e\cdot d$ and $3p_0$, respectively, where $d$ is the interlayer spacing between the electron and hole. $H_{\mathrm{2L}}$ produces a linear Stark shift for all $E_z$. $H_{\mathrm{3L}}$ gives eigenvalues $\lambda^{1,2}=\mathcal{E}_{d} \pm \sqrt{t{_h}^{2} + (p_{0}E_{z})^{2}}$. Near zero-field, the eigenvalues vary quadratically with $E_z$, approaching
$\lambda^{1,2}=\mathcal{E}_{d}\pm |t_h|$ at $E_{z}=0$, whereas at large
$E_z$ they recover a linear field dependence, $\lambda^{1,2}=\mathcal{E}_{d}\pm p_0E_z$. Thus, hole tunneling mixes the opposite dipolar configurations into a
quadrupolar state, explaining the nonlinear-to-linear Stark crossover~\cite{Yuobservation2023,Lianquadrupolar2023,Liquadrupolar2023,Xietransition2024,Deilmannquadrupolar2024,Maitymoire2026,Slobodkinquantum2020}. On the other hand, $H_{\mathrm{4L}}$  has no simple analytic expression for its eigenvalues. We therefore focus on their limiting behaviour. For large $E_{z}$, the four solutions take the form $\lambda^{1} = \mathcal{E}_d^{1} - p_{0}E_z$, $\lambda^{2} = \mathcal{E}_d^{2} + p_{0}E_z$, $\lambda^{3} = \mathcal{E}_d^{3} - p_{0}E_z$, and $\lambda^{4} = \mathcal{E}_d^{4} - 3p_{0}E_z$. The first dipole in fig.~\ref{fig1}(f) corresponds to the branch $\lambda^{2}$, while the last three dipoles shown in the same figure correspond to $\lambda^{4}$, $\lambda^{1}$, and $\lambda^{3}$, respectively. Hole tunneling hybridizes the oppositely oriented dipoles into quadrupolar branches at the field $E_z^{\star} = (\mathcal{E}_d^{1} - \mathcal{E}_d^{2})/2p_0$ where two configurations become degenerate. The zero-field ordering of the dipolar energies thus determines the field range over which the quadrupolar character evolves into dipolar character, establishing a subtle field-driven quadrupole-to-dipole transition.

\begin{figure*}[ht!]
    \centering
    \includegraphics[width=1.0\linewidth]{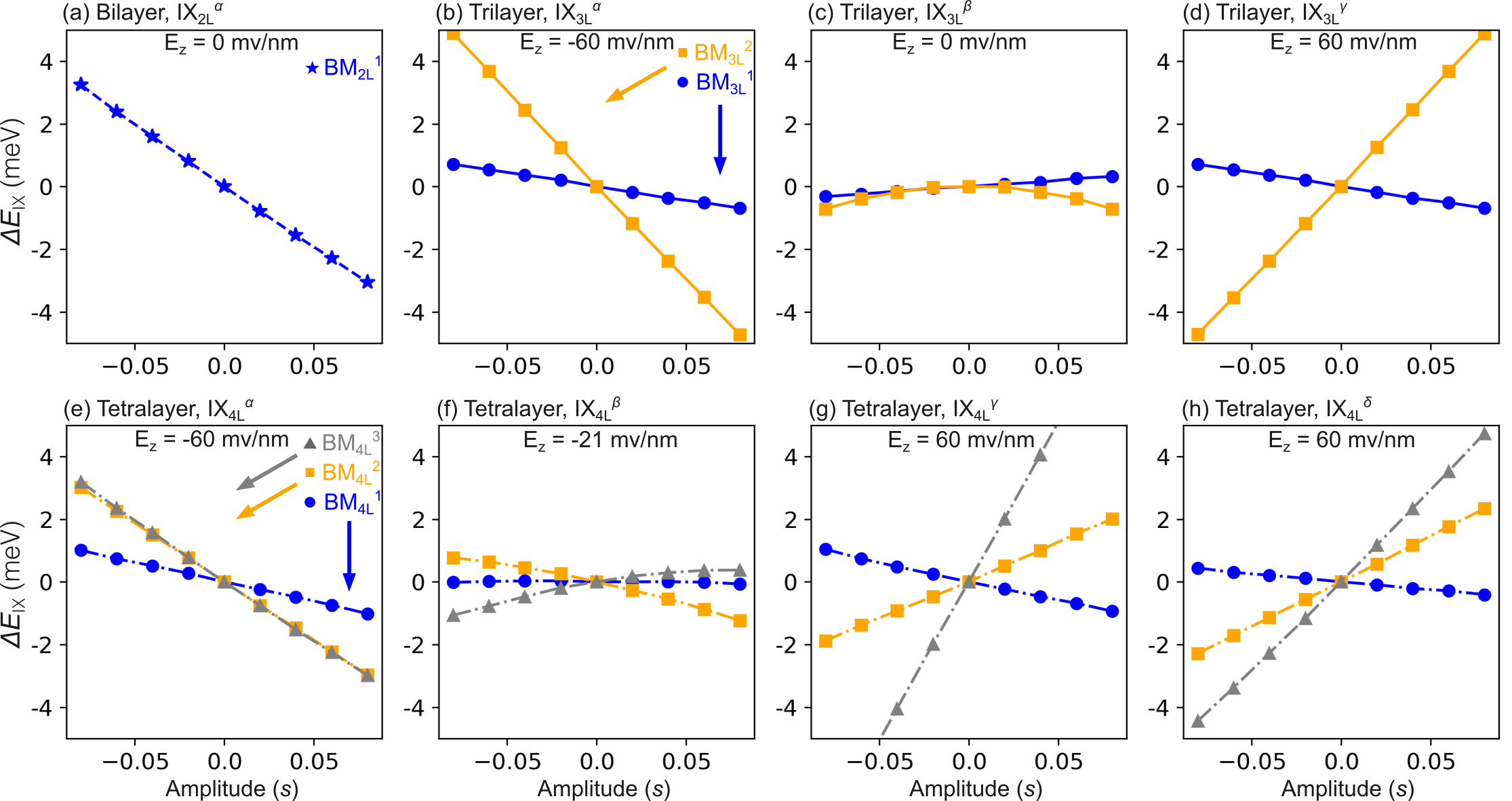}
    \caption{
    \textbf{Tunable exciton-breathing phonon coupling in multilayers of MoSe$_2$ and WSe$_2$}. (a) In the bilayer, the lowest-energy exciton $\mathrm{IX}_{2L}^{\alpha}$ couples linearly to the breathing mode $\mathrm{BM}_{2L}^{1}$. (b)-(d) In the trilayer, the higher-energy breathing mode $\mathrm{BM}_{3L}^{2}$ transitions from linear to quadratic coupling with the exciton multipolarity, whereas the lower-energy mode $\mathrm{BM}_{3L}^{1}$ remains linearly coupled across all fields. (e)-(g) The tetralayer shows a similar linear-to-quadratic-to-linear evolution. (h) highlights the brightest interlayer exciton-phonon coupling at $+60$~mV/nm, 30~meV above the lowest-energy exciton. Figs.~\ref{fig1} and \ref{fig2} show the internal structure of the excitons and phonons, respectively.
    }
    \label{fig3}
\end{figure*}

We compare our predictions with available experimental results. Experiments on MoSe$_{2}$/WSe$_{2}$ and WSe$_{2}$/MoSe$_{2}$/WSe$_{2}$ use magnetic-field-dependent and polarization-resolved photoluminescence spectra to identify singlet and triplet IXs. The measured singlet state is approximately 23 meV above the triplet state~\cite{Wanggiant2019,Xiebright2023}, consistent with our calculated value of 18 meV. Our calculated $t_{h}$ of $12 \pm 1\,\mathrm{meV}$ is also in agreement with the measured value of $16\,\mathrm{meV}$~\cite{Xiebright2023}. Although experiments have not yet probed tetralayers, the larger dipole moment of the $3p_{0}$ state we predict is consistent with giant dipole moments of up to 3.18 $e\cdot\mathrm{nm}$ reported for IXs in multilayer WS$_2$ and InSe heterostructures~\cite{Zhuobservation2025}.

\subsection{Breathing modes in multilayers of MoSe$_{2}$ and WSe$_{2}$} 
The electron and hole distributions of the IX resemble point charges on alternating MoSe$_2$ and WSe$_2$ layers. Phonons that modulate the interlayer distance therefore couple strongly to the IX. Such phonons correspond to the BMs, which involve the relative out-of-plane motion of adjacent layers. Fig.~\ref{fig2} shows the Brillouin-zone-center ($\Gamma$-point) BM energies and eigenvectors for a bilayer, a trilayer, and a tetralayer of alternating MoSe$_2$ and WSe$_2$, obtained by diagonalising the dynamical matrix (see Supplementary Information (SI), Sec.~B for the phonon dispersion). An $n$-layer system supports $(n-1)$ BMs, which map directly onto a linear chain of $n$ masses connected by $(n-1)$ springs~\cite{Tanshear2012, Lianglow2017, Luitemperature2014, Maitytemperature2018}. In all cases, BMs displace the layers along $z$ and transform as the totally symmetric (A-type) representation of the relevant point group. For a bilayer, the single BM at 0.82 THz describes antiphase motion, with the two layers moving toward and away from each other (see Fig.~\ref{fig2}(a)). For a trilayer, the lower-frequency mode at 0.53~THz leaves the central layer at rest while the outer layers move in antiphase with each other, and the higher-frequency mode at 1.04~THz has the outer layers moving in phase against
the central layer (see Fig.~\ref{fig2}(b)). For a tetralayer, the three modes at 0.44, 0.82 and 1.08~THz have more complex displacement patterns (see Fig.~\ref{fig2}(c)). Each mode modulates interlayer distances with a distinct pattern, and differently shifts the dipolar energies and electron and hole tunnelling described in Eqn.~\ref{avoided_crossing}. As a result, the exciton–phonon coupling depends on both the BM and the IX multipolarity. Importantly, these modes remain unchanged under an applied out-of-plane electric field, since the masses in the linear-chain model do not couple directly to $E_z$. Thus, the electric field provides direct control of the exciton–phonon coupling through IX multipolarity tuning.

\subsection{Electrically tunable exciton-phonon coupling} 
To quantify exciton-phonon coupling, we use a frozen-phonon approach. For each $\Gamma$-point BM, we construct a series of multilayer structures by displacing the atoms along the corresponding phonon eigenvector. We then solve the spinor BSE for each displaced structure to determine the exciton energy, $E_{\mathrm{IX}}(s)$, as a function of the displacement amplitude $s$. We fit $E_{\mathrm{IX}}(s)$ to
\begin{equation}
E_{\mathrm{IX}}(s)
= E_{\mathrm{IX}}(0)
+ \frac{g^\mu_{\mathrm{IX}}}{\ell_\mu}\,s
+ \frac{\lambda^\mu_{\mathrm{IX}}}{\ell_\mu^2}\,s^2,
\label{parabolic_exph}
\end{equation}
where $g^\mu_{\mathrm{IX}}$ and $\lambda^\mu_{\mathrm{IX}}$ are the linear and quadratic exciton-phonon coupling strengths, respectively, and $\ell_{\mu} = \sqrt{\hbar/2\omega_{\mu} m_{\mu}}$ is the zero-point amplitude of the mode, with $\omega_\mu$ and $m_\mu$ its angular frequency and reduced mass, respectively. Fig.~\ref{fig3} shows the exciton energy shifts as a function of phonon displacement in  multilayers under different out-of-plane electric fields. The lowest-energy IX in the bilayer is dipolar, as shown in Fig.~\ref{fig1}(d), and couples linearly to the only BM at 0.82~THz (Fig.~\ref{fig3}(a)). This coupling is independent of $E_z$. In contrast, the field-driven dipolar-to-quadrupolar transition in the trilayer, shown in Fig.~\ref{fig1}(b), \textit{qualitatively} changes the exciton-phonon coupling from linear to quadratic. Figs.~\ref{fig3}(b)-(d) demonstrate this change for $\mathrm{BM}^{2}_{3\mathrm{L}}$ at $1.04$~THz. However, $\mathrm{BM}^{1}_{3\mathrm{L}}$ at $0.53$~THz remains linear at all fields, while its coupling is about five times weaker than that of $\mathrm{BM}^{2}_{3\mathrm{L}}$ at large field. In the tetralayer, Figs.~\ref{fig3}(e)-(g) reveal that the more complex exciton multipolar structures lead to coexisting linear and quadratic coupling to multiple breathing phonons. Table~\ref{exph_params} summarizes the exciton-phonon coupling parameters corresponding to Fig.~\ref{fig3}, obtained from fits of the exciton energies to Eq.~\ref{parabolic_exph}. Our results establish that electrically tunable exciton multipolarity provides a systematic means of tuning the exciton-phonon coupling. This is the main result of this work.

We show the origin of mode-selective linear and quadratic exciton-phonon
coupling using the trilayer Hamiltonian in Eqn.~\ref{modelH} as an example. The interlayer spacing change shifts the dipolar energy by $\delta \mathcal{E}_{d}=\beta s$. The two BMs change the interlayer spacings differently: $\mathrm{BM}^{1}_{3\mathrm{L}}$ increases both spacings, whereas $\mathrm{BM}^{2}_{3\mathrm{L}}$ decreases one and
increases the other. At $E_z=0$, $H_{\mathrm{3L}}(s)$ becomes
\begin{equation}
  \begin{pmatrix}
    \mathcal{E}_d + \beta' s & -t_h \\
    -t_h & \mathcal{E}_d + \beta' s
  \end{pmatrix}
  \qquad \text{and} \qquad
  \begin{pmatrix}
    \mathcal{E}_d - \beta s & -t_h \\
    -t_h & \mathcal{E}_d + \beta s
  \end{pmatrix},
\end{equation}
with eigenvalues $\mathcal{E}_d + \beta' s \pm |t_h|$ and
$\mathcal{E}_d \mp \sqrt{t_h^{2}+\beta^{2}s^{2}}$, respectively. We approximate
$t_h$ to be independent of $s$ for $\mathrm{BM}^{2}_{3\mathrm{L}}$. The first mode shifts both configurations equally, giving linear coupling. The second mode detunes the two configurations, giving quadratic coupling.

\begin{table}[h!]
\centering
\begin{tabular}{c c c c c}
\hline\hline
Material & $\omega$ (THz) & $E_{z}$ (meV/nm) & $g^{\mu}_{\mathrm{IX}}$ (meV) & $\lambda^{\mu}_{\mathrm{IX}}$ (meV) \\
\hline
Bilayer   & 0.82 & -60 & -3.15 $\pm$ 0.07 & 0.00 \\
Trilayer   & 0.53 & -60 & -0.80 $\pm$ 0.01 & 0.00 \\
          &      &   0 & 0.36 $\pm$ 0.09 & 0.00 \\
          &      &   60 & -0.80 $\pm$ 0.01 & 0.00 \\
          & 1.04 & -60 & -4.38 $\pm$ 0.01 & 0.00 \\
          &      &   0 & 0.00 & -0.60 $\pm$ 0.01 \\
          &      &   60 & 4.38 $\pm$ 0.01 & 0.00 \\
Tetralayer   & 0.44 & -60 & -1.34 $\pm$ 0.01 & 0.00 \\
          &      &   -21 & 0.00 & -0.10 $\pm$ 0.02 \\
          &      &   60 & -1.30 $\pm$ 0.01 & 0.10 $\pm$ 0.02 \\
          & 0.82 & -60 & -2.99 $\pm$ 0.01 & 0.00 \\
          &      &   -21 & -1.00 $\pm$ 0.01 & -0.24 $\pm$ 0.01 \\
          &      &   60 & 1.94 $\pm$ 0.01 & 0.00 \\
          & 1.08 & -60 & -2.70 $\pm$ 0.01 & 0.00 \\
          &      &   -21 & 0.65 $\pm$ 0.01 & -0.27 $\pm$ 0.01 \\
          &      &   60 & 7.10 $\pm$ 0.01 & 0.00 \\
\hline\hline
\end{tabular}
\caption{Lowest IX-breathing-phonon coupling parameters extracted from fits of
$E_{\mathrm{IX}}(s)$ to Eq.~\ref{parabolic_exph}. Coefficients more than an order of magnitude smaller than the dominant one are set to zero.}
\label{exph_params}
\end{table}

\subsection{Coherent vs. squeezed phonons} 
Building on this insight, we reveal that photoexcitation drives coherent or squeezed phonons, depending on whether the BM couples linearly or quadratically to excited electrons. Photoexcitation with photon energies above the bandgap promotes electrons from the valence band to the conduction band. Ultrafast charge transfer then localizes electrons at the MoSe$_2$ conduction band minimum and holes at the WSe$_2$ valence band maximum~\cite{Duncanphotoinduced2025,Licoherent2023}. The resulting change in electron occupation drives phonons out of equilibrium through electron–phonon coupling. In particular, the lowest-energy IX couples to phonons through exciton–phonon coupling. We characterize photoinduced phonon states using the expectation values of displacement and momentum, $\langle \hat Q_{\mu}\rangle$, $\langle \hat P_\mu\rangle$, and their squares, $\langle \hat Q_{\mu}^{2}\rangle$, $\langle \hat P_\mu^{2}\rangle$. For linear coupling, these are 
\begin{equation}
\begin{aligned}
\langle\hat{Q}_\mu(t)\rangle 
&= -\frac{2\ell_\mu G_\mu^0}{\omega_\mu}\left(1-\cos\omega_\mu t\right),\\
\langle\hat{Q}_\mu^2(t)\rangle 
&= \ell_\mu^2 + \langle\hat{Q}_\mu(t)\rangle^2,\\[3pt]
\langle\hat{P}_\mu(t)\rangle
&= -\frac{\hbar G_\mu^0}{\omega_\mu\ell_\mu}\sin\omega_\mu t,\\
\langle\hat{P}_\mu^2(t)\rangle
&= \frac{\hbar^2}{4\ell_\mu^2} + \langle\hat{P}_\mu(t)\rangle^2,
\end{aligned}
\label{linear_elph}
\end{equation}
where $\hbar$ is the reduced Planck constant, $G_\mu^0 \propto n_{\rm exc}\,g^\mu_{\rm IX}$ with $n_{\rm exc}$ the photoexcited carrier density. In contrast, for quadratic coupling, these are
\begin{equation}
\begin{aligned}
\langle\hat{Q}_\mu(t)\rangle 
&= 0,\\
\langle\hat{Q}_\mu^2(t)\rangle 
& = \ell_\mu^2\left[1-\frac{2\omega_\mu\Lambda^{\mu\mu,0}}{\tilde\omega_\mu^2}\Big(1-\cos2\tilde\omega_\mu t\Big)\right],\\[3pt]
\langle\hat{P}_\mu(t)\rangle 
&= 0,\\
\langle\hat{P}_\mu^2(t)\rangle 
& = \frac{\hbar^2}{4\ell_\mu^2}\left[1+\frac{2\Lambda^{\mu\mu,0}}{\omega_\mu}\Big(1-\cos2\tilde\omega_\mu t\Big)\right], 
\end{aligned}
\label{quadratic_elph}
\end{equation}
where $\Lambda^{\mu\mu,0} \propto n_{\rm exc}\,
\lambda^{\mu\mu}_{\rm IX}$ is the quadratic coupling strength, and
$\tilde{\omega}_{\mu} = \sqrt{\omega_{\mu}^{2} + 4\Lambda^{\mu\mu,0}\omega_{\mu}}$
is the renormalized phonon frequency. See SI, Sec.~C for details of the derivation. In the linear-coupling case, the mean displacement is nonzero and oscillates at $\omega_\mu$, while the variance, $\langle\hat{Q}_\mu^2(t)\rangle-\langle\hat{Q}_\mu(t)\rangle ^2$ remains constant. These are signatures of a \textit{coherent} phonon state~\cite{Zeigertheory1992,Kuznetsovtheory1994}. In sharp contrast, the quadratic-coupling case maintains zero mean displacement, while the variances oscillate at $2\tilde{\omega}_\mu$. These are signatures of a \textit{squeezed} phonon state. When both linear and quadratic couplings are present, the resulting phonon state is both displaced and squeezed. See the SI movies for visual illustrations. In our calculations, we neglect anharmonic phonon damping~\cite{Emeiscoherent2025} and approximate the photoinduced electron population change by a Heaviside step function~\cite{Perfettotheory2024}. Neither approximation affects the signatures distinguishing coherent from squeezed phonons.

\begin{figure}
    \centering
    \includegraphics[width=\linewidth]{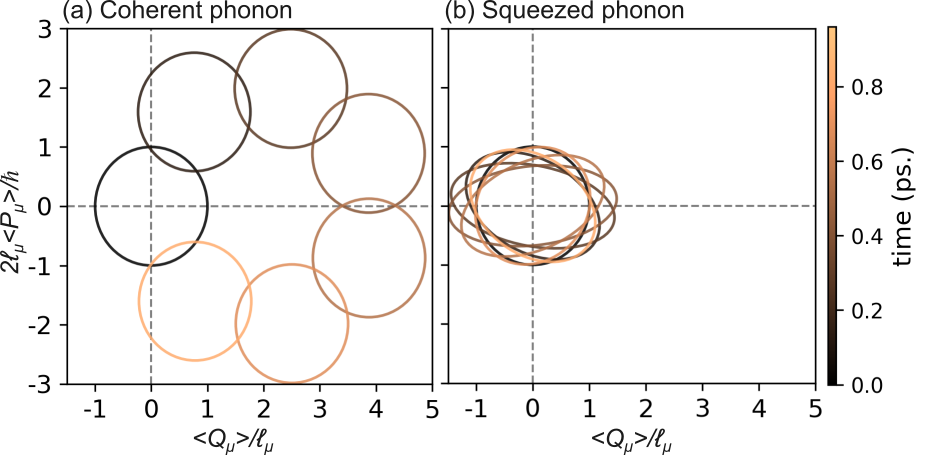}
    \caption{\textbf{Phase-space evolution of coherent and squeezed phonon states}. (a) The coherent phonon displaces without deforming. (b) The squeezed phonon deforms without displacing. $\langle Q_{\mu}\rangle$ and $\langle P_{\mu}\rangle$ are phonon displacement and momentum, with $\ell_{\mu}$ the zero-point amplitude.}
    \label{fig4}
\end{figure}

As an illustration, Figs.~\ref{fig4}(a) and (b) show the phase-space evolution of coherent and squeezed phonons for the 1.04\,THz trilayer BM. We use $g_{\mathrm{IX}}=-4.38$ meV and $\lambda_{\mathrm{IX}}=-0.6$ meV at $n_{\mathrm{exc}}=1$ to highlight the qualitative differences. Each ellipse represents the phonon state at time $t$, centered at the expectation values of the phonon coordinates and momenta, with its shape determined by the covariance matrix from Eqs.~\ref{linear_elph} and \ref{quadratic_elph}. The coherent phonon displaces in phase space over time without deforming (Fig.~\ref{fig4}(a)). In contrast, the squeezed phonon deforms over time without displacing (Fig.~\ref{fig4}(b)). We estimate displacement amplitudes and squeezing parameters for a typical pump-probe experiment, taking $n_{\mathrm{exc}}\sim3.7\times10^{-3}$ excitons per unit cell, corresponding to 1\% absorption of a 0.1~mJ/cm$^{2}$ pump fluence at 775~nm~\cite{Licoherent2023}. For the dipolar exciton at high $E_z$, the linear coupling gives equilibrium displacement of $\langle Q_\mu\rangle_{\text{eq}}\sim0.05$ pm for the 1.04 THz trilayer BM. The quadratic coupling squeezes the same mode at zero field, giving $|r|\sim1.0\times10^{-3}$ at the same excitation density. Here, $r= -\frac{1}{2}\ln\left[1-\frac{4\omega_\mu\Lambda^{\mu\mu,0}}{\tilde\omega_\mu^{2}}\right]$ quantifies the strength of the squeezing.

We compare our estimates with measurements in bismuth, a prototypical system for coherent
phonon generation. Experiments measure coherent $\mathrm{A_{1g}}$ phonon displacements of
$2.7\,\mathrm{pm}$~\cite{Johnsondirect2013} and $1.15\,\mathrm{pm}$~\cite{Teitelbaumdirect2018}
at pump fluences of $6$ and $2.5\,\mathrm{mJ/cm^{2}}$, respectively. Our estimate of
$0.05\,\mathrm{pm}$ at $0.1\,\mathrm{mJ/cm^{2}}$ is comparable when normalised by fluence.
Johnson \textit{et al.}~\cite{Johnsondirectly2009} measured mean-square atomic displacements
for specific planes in bismuth at $\sim1\,\mathrm{mJ/cm^{2}}$, reporting
$\Delta\langle(\hat{u}\cdot\hat{h})^2\rangle\approx5\times10^{-4}\,\text{\AA}^2$, a fractional
change $\delta\approx0.035$ of the equilibrium value
$\langle u^2\rangle_0=0.0143\,\text{\AA}^2$. Since $\lambda^{\mu\mu}_{\rm IX}<0$, the squeezing
resides in the momentum quadrature and we use magnitudes throughout. Using
$\langle u^2\rangle/\langle u^2\rangle_0 = e^{2|r|}$, this gives
$|r|=\tfrac12\ln(1+\delta)\approx2\times10^{-2}$ per $\mathrm{mJ/cm^{2}}$, comparable to our
predicted $|r|\sim1\times10^{-3}$ at $0.1\,\mathrm{mJ/cm^{2}}$. The bismuth measurement is at
$300\,\mathrm{K}$ and integrates over the Brillouin zone, whereas our estimate refers to a single zone-centre mode.

An important distinction arises in transparent materials, where light absorption is negligible. In this regime, coherent phonons can be generated through impulsive stimulated Raman scattering~\cite{Stevenscoherent2002}. First-order Raman scattering can excite all Raman-active modes, including the breathing and shear modes, with $\langle \hat{Q}_\mu \rangle \propto \sin(\omega_\mu t)$. Our mechanism instead selectively excites only the breathing modes, with $\langle \hat{Q}_\mu \rangle \propto \cos(\omega_\mu t)$~\cite{Zeigertheory1992}. Phonon squeezing originates from a second-order Raman process~\cite{Garrettvacuum1997}, which involves either a pair of phonons with equal and opposite momenta or a single phonon mode. Here, we focus specifically on single-mode squeezing.

Time-resolved X-ray diffraction directly reveals signatures of photoexcited coherent and squeezed phonons in bulk crystals, through intensity modulations from mean atomic displacements~\cite{Johnsonnanoscale2008} and Debye–Waller factor changes~\cite{Johnsondirectly2009} respectively, with transient optical measurements offering a complementary, indirect route to the same signatures~\cite{Zeigertheory1992,Garrettvacuum1997}. Recent ultrafast electron diffraction~\cite{Duncanphotoinduced2025} and transient reflectivity~\cite{Licoherent2023} experiments have observed coherent phonons in MoSe$_2$/WSe$_2$ bilayers, demonstrating that these signatures persist in van der Waals heterostructures.

In summary, our atomistic calculations establish exciton multipolarity as a design principle for generating photoinduced coherent and squeezed breathing phonons in multilayers of MoSe$_2$ and WSe$_2$. Moreover, small twist angles provide a direct means to further tune exciton–phonon coupling~\cite{Uzundalmoire2026}, opening new avenues for exploring ultrafast phenomena in moiré materials. The electric-field-switchable coherent-to-squeezed transition offers a solid-state counterpart to squeezed-state protocols in quantum optics and optomechanics—such as those already used to suppress quantum noise below the standard limit in gravitational-wave interferometers like LIGO~\cite{Wenxuansqueezing2024}—pointing toward THz-frequency, chip-based quantum sensing built from an electrically controlled phonon.

\section*{Author contributions}
I.~M. conceived the project. I.~M. and A.~R. developed strategies for experimental observation. A.~A.~M. provided computing resources. I.~M. developed the algorithms, carried out the simulations, and wrote the first draft of the manuscript. All authors discussed the results and contributed to the manuscript.

\section*{Ethics declaration}
The authors declare no competing interests.

\section*{Code availability}
Structure construction, atomic relaxations, electronic structure, and exciton calculations presented in this work were performed using publicly available codes. PyMEX is available at \url{https://github.com/imaitygit/PyMEX}.

\section*{Acknowledgements}
This work used the ARCHER2 UK National Supercomputing Service (https://www.archer2.ac.uk)~\cite{BeckettARCHER22024} through our membership of the UKCP consortium (EP/X035891/1), funded by EPSRC. 

\clearpage
\newpage

\section{Methods}

\noindent \textbf{Atomic structure.} Atomic relaxations were carried out using classical interatomic potentials in \textsc{LAMMPS}~\cite{lammps, Thompsonlammps2022}. Intralayer interactions were modeled using the Stillinger–Weber potential~\cite{Zhouhandbook2017}, while interlayer interactions were described by the Kolmogorov–Crespi potential~\cite{Naikkolmogorov2019}. For the latter, interactions were included only between neighboring layers, as these dominate over contributions from next-nearest-neighbor layers~\cite{Maitytemeperature2018}. For all multilayer structures, the in-plane lattice constant was set to 3.32 $\mathrm{\AA}$, while the interlayer separation between neighboring metal-atom layers was found to be 6.44 $\mathrm{\AA}$. 

\noindent \textbf{Electronic structure.} We calculated ground-state electronic structures within density functional theory (DFT), as implemented in Quantum ESPRESSO~\cite{Giannozziquantum2009,Giannozziadvanced2017,Giannozziquantum2020,quantumespresso}. Fully relativistic optimized norm-conserving Vanderbilt pseudopotentials generated with the ONCVPSP code~\cite{Hamanoptimized2013} were combined with the Perdew-Burke-Ernzerhof (PBE) generalized gradient approximation~\cite{Perdewgeneralized1996} for the exchange-correlation functional. We used plane-wave energy cutoffs of 100~Ry and 400~Ry for the wavefunctions and charge density, respectively. An 18~\AA\ vacuum spacing separated periodic images along the out-of-plane direction. Brillouin-zone sampling used a $15\times15\times1$ $k$-point grid for all calculations. Self-consistent-field calculations reached an energy threshold of $10^{-14}$~Ry, with spin-orbit coupling included throughout. The atomic structures described above were used directly in the DFT and phonon calculations, without further relaxation.

Spinor Wannier functions were generated via the one-shot projection method~\cite{Marzarimaximally2012} as implemented in the \textsc{Wannier90} code~\cite{Pizziwannier2020}, using spinor $d$ and $p$ orbitals as initial projections. Specifically, Kohn-Sham spinor wavefunctions were projected onto atom-centered $d$ ($d_{xy}$, $d_{yz}$, $d_{zx}$, $d_{x^2-y^2}$, $d_{z^2}$) orbitals on every W and Mo atom and $p$ ($p_x$, $p_y$, $p_z$) orbitals on every Se atom, with each orbital assigned both spin channels along the $z$-spin quantization axis, followed by orthogonalization and disentanglement~\cite{Souzamaximally2001}. We used minimal-distance replica selection for Fourier interpolation of the Wannier Hamiltonian~\cite{Maityorigin2026}.

\noindent \textbf{Phonon structure.} We calculated phonons using the finite-displacement method as implemented in PHONOPY~\cite{Togofirst2015}, interfaced with LAMMPS for force calculations using the interatomic potentials described above. A $6\times6\times1$ supercell was used to generate atomic displacements. The resulting forces were used to construct the force-constant matrix, from which we obtained and diagonalized the dynamical matrix.

\noindent \textbf{Exciton structure.} We employed PyMEX to construct and diagonalise the BSE Hamiltonian~\cite{Maityatomistic2025,Maitymoire2026} and compute optical conductivities using spinor Wannier functions as basis, with diagonalisation accelerated by the Eigenvalue soLvers for Petaflop Applications (\textsc{Elpa}) library~\cite{Auckenthalerparallel2011,Marekelpa2014}. We used the transfer-matrix approach to capture screened Coulomb interactions in multilayer structures~\cite{Maitymoire2026}. We used 2D polarizabilities of $\chi_{\mathrm{MoSe_2}}=103.36$~\AA\ and $\chi_{\mathrm{WSe_2}}=90.18$~\AA, obtained from DFT~\cite{Berkelbachtheory2013}. We applied a vertical electric field $E_{z}$ by adding the term $-eE_{z}\sum_{i} z_i |i\rangle \langle i|$ to the electronic Hamiltonian, where $z_i$ is the $z$-coordinate of the $i$-th Wannier function and $e$ is the electron charge. For each multilayer structures, we performed spinor-BSE calculations at 51 electric-field values and interpolated the results to construct a heatmap of the field-dependent optical conductivity. We used a $66\times66\times1$ $k$-point grid and included 1, 2, and 2 valence bands and 2, 2, and 4 conduction bands in the BSE Hamiltonian for the bilayer, trilayer, and tetralayer systems, respectively. The SI, Sec.~D, provides details of the convergence and justification for the chosen number of bands. 

The BSE Hamiltonian is expressed in terms of blocks in the two-component spinor basis as,
\begin{equation}
H_{\text{BSE}}^{\text{block}} =  ( \epsilon_{s_{1}} - \epsilon_{s_{2}})\delta_{s_{1},s_{3}}\delta_{s_{2},s_{4}} - W_{s_{1}s_{3}s_{2}s_{4}} + V_{s_{1}s_{2}s_{3}s_{4}}
\end{equation}
Here, $s$ denotes the spin index, $\epsilon_s$ denotes the energy eigenvalues, $W$ denotes the screened Coulomb potential, and $V$ denotes the bare Coulomb potential~\cite{Marsilispinorial2021}. This is a $4\times4$ matrix, and we use this approach to solve the BSE. For weaker spin-orbit coupling, the full Hamiltonian decouples into spin-triplet and spin-singlet solutions~\cite{Rohlfingelectron2000}. The spin-orbit splitting separates the solutions when the exchange contribution is negligible.

\noindent \textbf{Exciton-phonon coupling.} Exciton-phonon couplings were obtained from frozen-phonon calculations. For
each $\Gamma$-point breathing mode of the equilibrium multilayer structure, we
constructed a series of displaced configurations
\begin{equation}
  \begin{aligned}
    \mathbf{R}_{\kappa}
      &= \mathbf{R}^{0}_{\kappa} + s\,\hat{\mathbf{u}}_{\mu,\kappa}, \\[4pt]
    \hat{\mathbf{u}}_{\mu,\kappa\alpha}
      &= \frac{1}{N_{\mu}}\,\frac{e^{\mu}_{\kappa\alpha}}{\sqrt{M_{\kappa}}}, \\[4pt]
    N_{\mu}
      &= \Biggl[\,\sum_{\kappa\alpha}
         \frac{|e^{\mu}_{\kappa\alpha}|^{2}}{M_{\kappa}}\Biggr]^{1/2},
  \end{aligned}
  \label{eq:frozen}
\end{equation}
where the sum over $\kappa$ runs over all atoms of the multilayer,
$e^{\mu}_{\kappa\alpha}$ is the mass-weighted phonon eigenvector normalized as
$\sum_{\kappa\alpha}|e^{\mu}_{\kappa\alpha}|^{2}=1$, and $M_{\kappa}$ is the
mass of atom $\kappa$. The factor $N_{\mu}$ normalizes the Cartesian
displacement pattern to unit norm, so that the amplitude $s$ is the total
Cartesian amplitude in \AA, and the reduced mass of the mode follows as
$m_{\mu}=N_{\mu}^{-2}$. For each mode we used nine amplitudes spanning $s$ from
$-0.08$ to $+0.08$~\AA{} in steps of $0.02$~\AA, and solved
the spinor BSE at each amplitude to obtain $E_{\mathrm{IX}}(s)$. The energies
were fit by least squares to the second-order polynomial of
Eq.~\eqref{parabolic_exph}. Uncertainties are the standard errors of the fitted
coefficients. See SI, Sec.~E for examples of the fits. The linear and quadratic coefficients were obtained from a single fit. If one coefficient exceeded the other by more than an order of magnitude, only the dominant coefficient was reported.

\bibliography{breathing}
\end{document}